# Nonlinear electric field effect on perpendicular magnetic anisotropy in Fe/MgO interfaces

Qingyi Xiang[1,2,5], Zhenchao Wen[3,4], Hiroaki Sukegawa[2], Shinya Kasai[2], Takeshi Seki[3,4], Takahide Kubota[3,4], Koki Takanashi[3,4] and Seiji Mitani[1,2]

[1] Graduate School of Pure and Applied Sciences, University of Tsukuba, Tsukuba 305-8577, Japan
[2] National Institute for Materials Science, Tsukuba, 305-0047, Japan
[3] Center for Spintronics Research Network, Tohoku University, Sendai 980-8577, Japan
[4] Institute for Materials Research, Tohoku University, Sendai 980-8577, Japan

E-mail: XIANG.Qingyi@nims.go.jp

Abstract

The electric field effect on magnetic anisotropy was studied in an ultrathin Fe(001) monocrystalline layer sandwiched between Cr buffer and MgO tunnel barrier layers, mainly through post-annealing temperature and measurement temperature dependences. A large coefficient of the electric field effect of more than 200 fJ/Vm was observed in the negative range of electric field, as well as an areal energy density of perpendicular magnetic anisotropy (PMA) of around 600 μJ/m$^2$. More interestingly, nonlinear behavior, giving rise to a local minimum around +100 mV/nm, was observed in the electric field dependence of magnetic anisotropy, being independent of the post-annealing and measurement temperatures. The insensitivity to both the interface conditions and the temperature of the system suggests that the nonlinear behavior is attributed to an intrinsic origin such as an inherent electronic structure in the Fe/MgO interface. The present study can contribute to the progress in theoretical studies, such as *ab initio* calculations, on the mechanism of the electric field effect on PMA.

[5] Corresponding author

## 1. Introduction

The electric field effect on magnetic anisotropy, which is often called "voltage-control of magnetic anisotropy (VCMA)", in ferromagnetic metal layers attracts much interest in recent years [1–24]. This technology can help to realize low-power magnetization switching in devices, which is a key to next generation magnetic random access memories (MRAMs). In fact, by using the VCMA, direct manipulation of magnetization by voltage [25–27] and assistance to spin transfer torque switching [28,29] have been demonstrated.

Cr-buffered Fe/MgO heterostructures are of particular importance in perpendicular magnetic anisotropy (PMA) and relevant studies, and Nozaki *et al*. have recently achieved a remarkable progress in the VCMA study using magnetic tunnel junctions (MTJs) with the Cr/Fe/MgO structure [23]. First, a very large interface PMA was predicted for the Cr/Fe/MgO structure by *ab initio* calculations [30,31], followed by an experimental demonstration of PMA of ~1.4 MJ/$m^3$ (~1000 μJ/$m^2$ for areal energy density) in previous studies [32,33]. Then, Nozaki *et al*. successfully reproduced such a large PMA in their MTJs including the Cr/Fe/MgO structure, so that its VCMA was examined. The VCMA coefficient obtained for the PMA reached 290 fJ/Vm [23].

The mechanism of the large VCMA in Cr/Fe/MgO has not been well understood,

although *ab initio* studies have described the effect of charge accumulation and depletion at the interface that affects the spin-dependent screening and electron's occupancy of 3*d* orbitals [32,34]. Furthermore, an open question is the nonlinear behavior observed unexpectedly in the electric field dependence of PMA energy density [23]. Since it appears to be a unique feature for the VCMA coefficient more than 200 fJ/Vm, the nonlinear behavior could be a key to develop further large VCMA. At the same time, one may wonder if the nonlinearity can be associated with possible sample-to-sample variation, since no theoretical explanation has been provided. As Cr diffusion into the Fe layer (also into the Fe/MgO interface region) and possible interface contamination with carbon and/or oxygen are somehow discussed in Ref. 23, it may be difficult to obtain the well-defined Fe/MgO interface repeatedly and systematically.

In this work, we examined the VCMA in Cr/Fe/MgO under different conditions, i.e., temperature for annealing the structure and temperature of the VCMA measurement, to confirm the presence of the nonlinear VCMA behavior occurring characteristically in the Cr/Fe/MgO with a large VCMA coefficient. The nonlinear VCMA that was obtained together with a large areal PMA energy of ~600 $\mu$J/m$^2$ and a large VCMA coefficient of more than 200 fJ/Vm gave rise to a local minimum at around 100 mV/nm in the VCMA curve, being independent of both the post-annealing temperature (i.e., interface quality) and the measurement temperatures. The results insensitive to the interface quality and the measurement conditions suggest that the nonlinear VCMA has an intrinsic origin such as a basic feature of the interface electronic structure.

## 2. Experimental procedures

Figure 1 shows a schematic design of the MTJ used to examine the VCMA of Cr/Fe/MgO in this study. A fully epitaxial stack of MgO (5 nm)/Cr (30 nm)/Fe (0.7 nm)/MgO (2.2 nm)/Fe (2 nm)/Ru (15 nm) was deposited on a MgO (001) substrate by electron beam evaporation. The bottom 0.7 nm Fe layer is the free layer to respond the voltage effect with different PMA energy density, while the 2 nm top Fe layer with in-plane magnetization is the reference layer to quantitatively evaluate the VCMA effect. Here, the thickness of the bottom (free) Fe layer $t_{Fe}$ corresponds to five monolayers of bcc-Fe(001). At the beginning, the substrate was annealed at 800°C to clean its surface, followed by the deposition of a 5 nm MgO seed layer at 450°C. A Cr buffer layer was deposited at 150°C, and then was post-annealed at 800°C to obtain a flat Cr(001) surface. This crucial post-annealing temperature is critical to obtain large PMA for an ultrathin Fe layer deposited on the Cr buffer. The conditions used for the ultrathin Fe were 150°C for growth and 250°C for post-annealing to improve the surface flatness. Then, the MgO barrier layer was deposited at 150°C, followed with post-annealing at different temperatures (325°C, 350°C, 375°C, and 400°C) to modify the Fe/MgO interface conditions. The reference Fe layer was deposited at 150°C without post-annealing. Finally, a Ru capping layer was sputter-deposited at room temperature (RT). Through the growth process, surface structures and epitaxial growth were monitored by reflection high-energy electron diffraction (RHEED). The MTJ multilayer stacks prepared were patterned into a 5×10 μm$^2$ elliptical shape by using photo-lithography, Ar ion-beam milling and lift-off processes. The tunnel magnetoresistance (TMR) ratio was measured using the physical property measurement system (PPMS) under an in-plane external magnetic field at RT and low temperatures. The positive sign of applied voltage was defined as the polarity corresponding to the current flow from the top to bottom Fe electrode. The typical resistance-area product (*RA*) of the MTJs was the order of $10^4$ $\Omega\mu m^2$. Separately, the magnetization curve of the ultrathin Fe layer was characterized by using a vibrating sample magnetometer to determine the saturation magnetization ($M_s$).

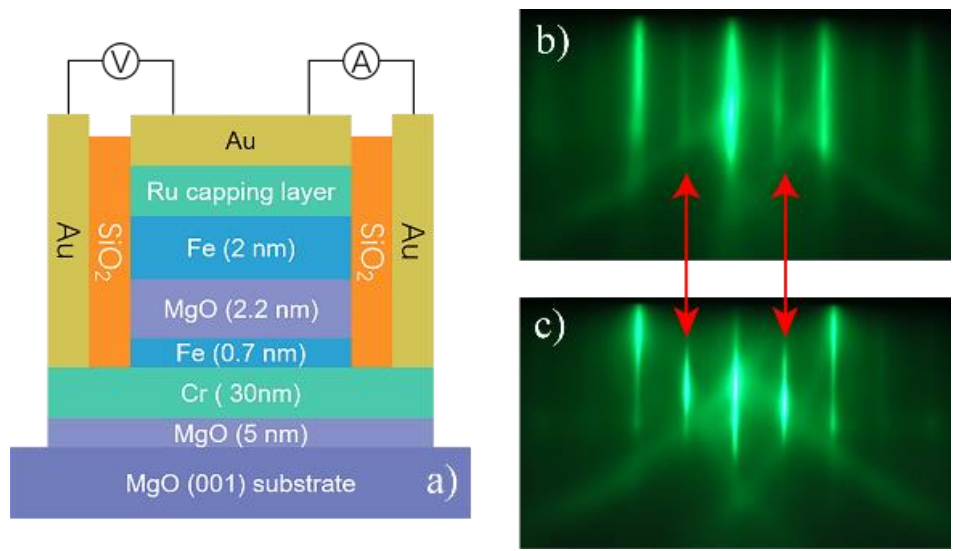

**Figure 1.** (a) Schematic illustation of epitaxial magentic tunnel junctions in this study, (b)(c) RHEED patterns taken for the surfaces of (b) ultrathin Fe and (c) Cr buffer layers. The incident electron beams are along the [100] azimuth of MgO(001) substrate. Sub-streaks indicated by red arrows correspond to a c(2×2) surface structure.

## 3. Results and discussion

Figures 1(b) and 1(c) show the RHEED patterns of the ultrathin Fe (0.7 nm) and the Cr buffer layers with the incident electron beam parallel to the [100] azimuth of the MgO(001) substrate. Formation of c(2×2) surface structure is observed for both the Cr and Fe surfaces, as the additional streaks pointed by red arrows. Such a surface structure may improve the surface flatness and the magnitude of PMA of the ultrathin Fe layer [31]. In addition, it is noted that the absence of the c(2×2) structure for Fe was reported in Ref. 23, in contrast to the present study.

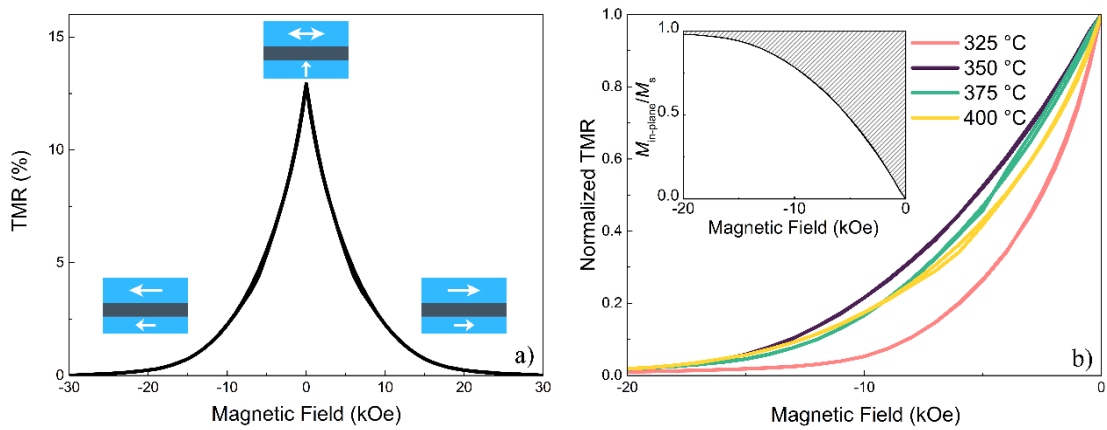

**Figure 2.** (a) A typical TMR curve for the MTJ with annealing temperature of 400°C, (b) normalized TMR curves for the MTJs with different annealing temperatures. The inset is an example of the in-plane component of magnetization (annealing tempaerture of 350°C), where the shadow area corresponds to the PMA energy density.

Figure 2(a) shows a full TMR curve of a prepared MTJ annealed at 400 °C in the in-plane magnetic field ($H_{ex}$). Due to the small shape magnetic anisotropy energy of the ultrathin Fe layer and the sufficiently large interface PMA induced at the Fe/MgO interface, the easy axis of the ultrathin Fe layer is aligned perpendicular to the film plane; meanwhile, the top Fe layer has an easy axis parallel to the film plane. This orthogonal-easy-axis design enables us to electrically detect the rotation of the magnetization in the ultrathin Fe layer by applying an $H_{ex}$ [11]. As described in Fig. 2(a), the two Fe layers take the orthogonal magnetization configuration at $H_{ex} = 0$, which brings about a high tunnel resistance state of the MTJ. By applying an $H_{ex}$, the magnetization of the top Fe layer (reference layer) immediately turns parallel to the $H_{ex}$ due to its in-plane magnetic anisotropy and small in-plane coercivity (<a few Oe), i.e., the magnetization of the top layer is almost always parallel to the $H_{ex}$ direction. On the other hand, the magnetization of the bottom Fe layer (free layer) gradually tilts and finally becomes parallel to the $H_{ex}$ direction when the $H_{ex}$ reaches the anisotropy field of the bottom Fe layer ($H_k$). Therefore, the TMR curve reflects the magnetization process of the free Fe layer, i.e., the tunnel resistance takes the maximum at $H_{ex} = 0$ (orthogonal magnetization configuration) and gradually decreases down to the minimum with the increase of $|H_{ex}|$ (parallel magnetization configuration). The TMR ratio in Fig. 2(a) corresponds to a half of the whole TMR change between parallel and antiparallel magnetization configurations.

Figure 2(b) shows the normalized TMR curves in the negative $H_{ex}$ region for MTJs annealed at different temperatures. The TMR ratio is normalized by using the maximum (at $H_{ex} = 0$) and minimum (at $H_{ex} > H_k$) values, respectively. In the TMR curves, one can clearly see that the saturation behavior changes with the annealing temperature. This means that $H_k$ strongly depends on the annealing temperature, indicating that the annealing process governs the Fe/MgO interface conditions that determine the PMA characteristics.

From the normalized TMR curves, we can evaluate the effective PMA energy density ($K_{eff}$), including the contribution of the shape magnetic anisotropy, as follows: the tunnel resistance is given by the relative angle between the magnetizations of the free and reference magnetic layers. In the sample design, the maximum resistance occurs at the 90° configurations ($R_{90}$ at $H_{ex}$ = 0), while the minimum resistance does in the parallel configuration ($R_p$ at $H_{ex}$ = $H_k$). The resistance at a given relative angle $\theta$ is expressed as [23]:

$$R(\theta) = \frac{R_{90} R_p}{R_p + (R_{90} - R_p) \cdot \cos\theta} \qquad (1)$$

Since the magnetization direction of the top Fe layer is considered to be parallel to $H_{ex}$ (i.e., in-pane direction), the ratio of the in-plane component of the magnetization $M_{\text{in-plane}}$ to its saturation magnetization $M_s$ in the bottom ultrathin Fe layer can be determined as:

$$\frac{M_{in-plane}}{M_s} = \cos\theta = \frac{R_{90} - R(\theta)}{R(\theta)} \frac{R_p}{R_{90} - R_p} \qquad (2)$$

The $M_{\text{in-plane}}/M_s$ calculated from the observed TMR curve of the MTJ with post-annealing at 350°C is shown in the inset of Fig. 2(b), as an example. The $K_{\text{eff}}$ is finally obtained by calculating the product of the area of shadow region $A_{\text{in-plane}}$ and the $M_s$.

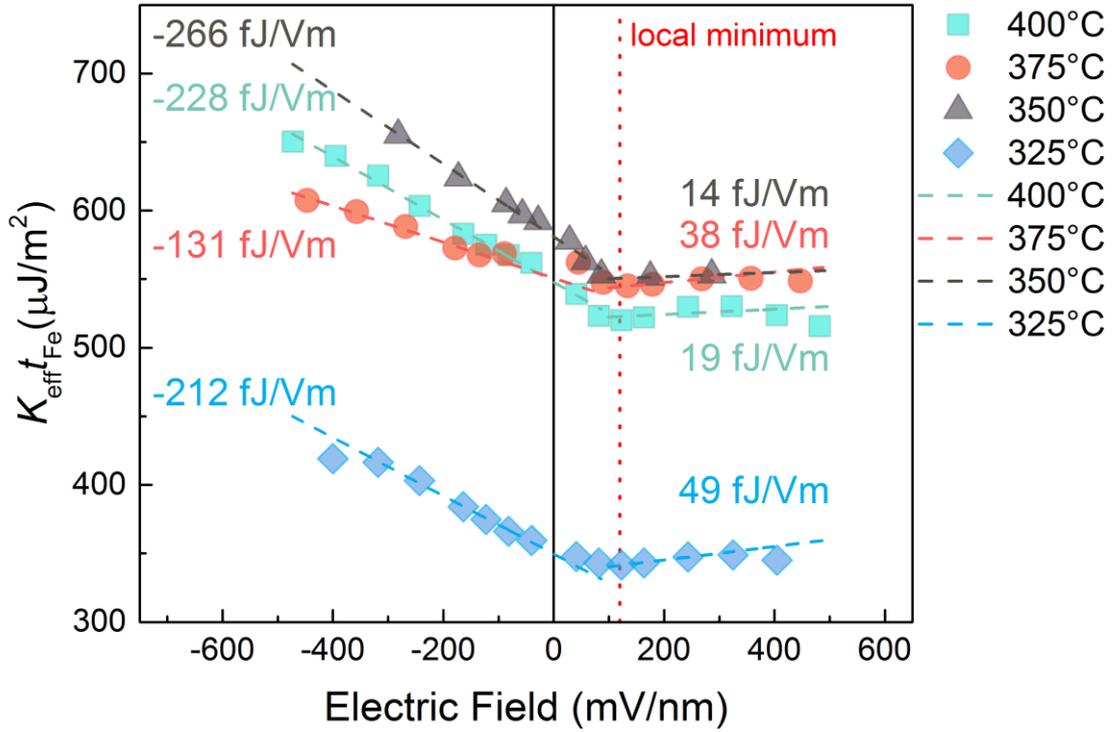

**Figure 3.** $K_{eff}\,t_{Fe}$ as a function of applied electric field $E$ at RT for MTJs annealed at different temperatures. The dash line is the linear fitting. The dotted line indicates the position of the local minimum.

Figure 3 summarizes the areal PMA energy densities ($K_{eff}\,t_{Fe}$) as a function of the external electric field ($E$) measured at RT for MTJs with different annealing temperatures, where $t_{Fe}$ = 0.7 nm and $E$ is defined as the applied voltage divided by the thickness of the MgO barrier (=2.2 nm). The applied voltage range of -1000 to +1000 mV corresponds to the $E$ range of -450 to +450 mV/nm. For all the MTJs, the $E$ dependences of $K_{eff}\,t_{Fe}$ have a local minimum at around +100 mV/nm, which is indicated by the red dotted line in Fig. 3. It is also found that $K_{eff}\,t_{Fe}$ varies almost linearly with $E$ below and above the local minima. The linear fitting results were plotted for each MTJ (dashed line). The slope below the local minimum positions were in the range of -133 to -266 fJ/Vm. The values of the slopes are the so-called

VCMA coefficient, and those observed in the present Fe/MgO interfaces are consistent with that reported in Ref. 23. In the range of $E$ above the local minima, the VCMA coefficients are much smaller than those below the local minima. Furthermore, it is a new finding in the present study that the local minimum position around +100 mV/nm is independent of the annealing temperature. This suggests that the nonlinear behavior is insensitive to the interface conditions, while the PMA energies and VCMA coefficients are very likely to depend on the interface conditions.

The appearance of the local minimum always observed at around +100 mV/nm despite the variation in the annealing temperature would be specific to the VCMA characteristics of Cr/Fe (0.7 nm)/MgO structures. To explore it further, the VCMA characteristics for the MTJ annealed at 400 °C was evaluated at low measurement temperatures. Figure 4 shows the $K_{eff}$ $t_{Fe}$ as a function of $E$ at 10, 100, 200 and 300 K, respectively. It is clearly seen that the $K_{eff} t_{Fe}$ increases with decreasing the measurement temperature. Interestingly, the local minimum positions (~ +100 mV/nm) are independent of the temperatures, while the nonlinearity becomes significant at lower temperatures. In addition, there might be a few fine structures, as implied by the faint peak at around -50 mV/nm, in the $E$ dependence of $K_{eff}$ $t_{Fe}$ at low temperatures.

As shown in Figs. 3 and 4, the nonlinear behavior, particularly the local minimum at around +100 mV/nm, has been found to be independent of both the annealing and measurement temperatures. The former temperature is likely to influence the interface conditions, and indeed the PMA and the VCMA coefficient vary with the annealing temperature. The latter temperature may be related with possible extrinsic effects such as an epitaxial strain induced in the MTJ and some kinds of artifacts. Thus, the observed insensitivity of the minimum position suggests that the nonlinear behavior is attributed to an intrinsic origin such as a basic feature of the electronic structure at the Fe/MgO interface. In fact, interface resonant states (IRSs) are formed in the minority spin band of the Fe/MgO system, and the IRSs may affect the transport properties in the Fe/MgO-based MTJs, as proposed by Belashchenko *et al*. [35]. The effect of IRSs on the VCMA at the Fe/MgO interface was also studied by means of *ab initio* calculations [36]. Thus, the results obtained are expected to contribute to the progress in theoretical studies, particularly in *ab initio* calculations [36-41], on the mechanism of VCMA.

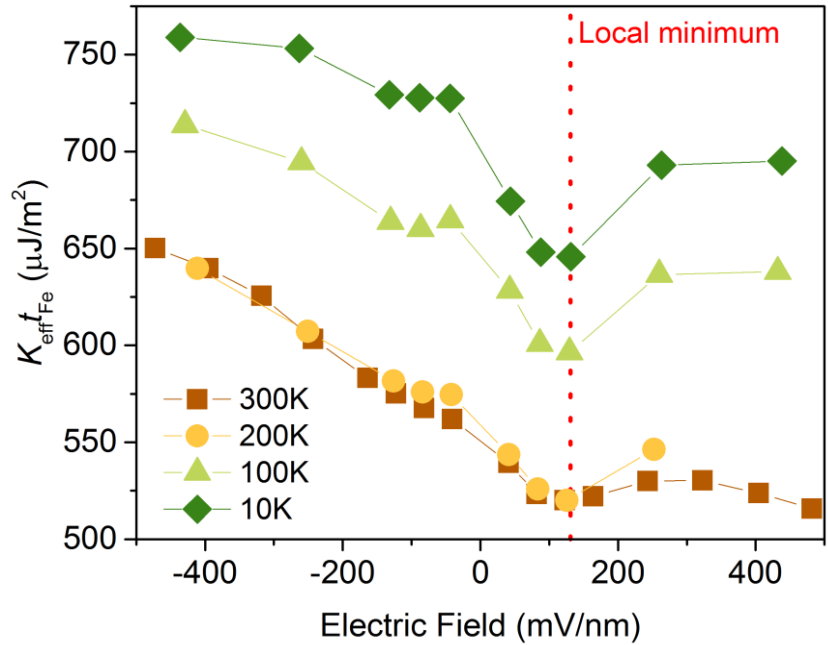

**Figure. 4** $K_{\text{eff}} t_{\text{Fe}}$ as a function of applied electric field $E$ at different measurement temperatures. The annealing temperature of MTJ is 400 °C.

## 4. Conclusions

The nonlinear VCMA characteristics in orthogonally-magnetized MTJs with a Cr/Fe(0.7nm)/MgO structure was studied by evaluating post-annealing and measurement temperature dependences. A large VCMA coefficient of more than 200 fJ/Vm and a large areal PMA energy density of around 600 $\mu J/m^2$ at RT were obtained at the 0.7-nm Fe/MgO interfaces. More interestingly, regardless of the post-annealing and measurement temperatures, a clear local minimum around +100 mV/nm was observed in the electric field dependence of magnetic anisotropy. The present results imply that the origin of the local minimum is attributed to an inherent electronic structure in the Cr/Fe/MgO.

**Acknowledgements**

This study was partly supported by the ImPACT program of the Council for Science, Technology and Innovation (Cabinet Office, Government of Japan) and JSPS KAKENHI Grant Number 16H06332. A part of this work was performed under the Inter-University Cooperative Research Program of Institute for Materials Research, Tohoku University and the JSPS-EPSRC-DFG Core-to-Core Program. Q.X. acknowledges National Institute for Materials Science for the provision of a NIMS Junior Research Assistantship.